\documentclass[12pt]{iopart}

\usepackage{iopams}  
\usepackage{amsmath}
\usepackage{graphicx}
\usepackage{booktabs}
\usepackage{amsmath, amsthm, amssymb, amsfonts}
\usepackage{thmtools}
\usepackage{graphicx}
\usepackage{setspace}
\usepackage{geometry}
\usepackage{float}
\usepackage{hyperref}
\usepackage{physics}
\usepackage[utf8]{inputenc}
\usepackage[english]{babel}
\usepackage{framed}
\usepackage[dvipsnames]{xcolor}
\usepackage{tcolorbox}
\usepackage{marvosym}
\usepackage{diagbox}
\usepackage{graphicx}
\usepackage{enumitem}

\usepackage{graphicx}
\usepackage{subcaption}

\DeclareMathOperator{\leakIR}{leak_{IR}}
\DeclareMathOperator{\leakEV}{leak_{EV}}
\colorlet{LightGray}{White!90!Periwinkle}
\colorlet{LightOrange}{Orange!15}
\colorlet{LightGreen}{Green!15}
\usepackage{booktabs}
\usepackage{diagbox}
\usepackage[resetlabels,labeled]{multibib}
\newcites{Math}{Math Readings}

\begin{document}

\title[Performance of Cascade and LDPC-codes on Industrial QKD systems]{Performance of Cascade and LDPC-codes for Information Reconciliation on Industrial Quantum Key Distribution Systems}

\author{Ronny Mueller$^1$,  Claudia De Lazzari$^2$, Fernando Chirici$^2$, Ilaria Vagniluca$^2$, Leif Katsuo Oxenl\o we$^1$, S\o ren Forchhammer$^1$, Alessandro Zavatta$^{2,3}$, Davide Bacco$^{2,4}$}

\address{$^1$ Department of Electrical and Photonics Engineering, Technical University of Denmark, Lyngby, Denmark}
\address{$^2$QTI S.r.l., Florence, Italy}
\address{$^3$National Institute of Optics, CNR-INO, Florence, Italy}
\address{$^4$ Department of Physics and Astronomy, University of Florence, Sesto Fiorentino, Italy}

\ead{ronmu@dtu.dk}

\vspace{10pt}

\begin{abstract}
Information Reconciliation is a critical component of Quantum Key Distribution, ensuring that mismatches between Alice's and Bob's keys are corrected. In this study, we analyze, simulate, optimize, and compare the performance of two prevalent algorithms used for Information Reconciliation: Cascade and LDPC codes in combination with the Blind protocol. We focus on their applicability in practical and industrial settings, operating in realistic and application-close conditions. The results are further validated through evaluation on a live industrial QKD system.
\end{abstract}

\newcommand{\qedwhite}{\hfill \ensuremath{\Box}}
%
%
%
%

\section{Introduction}

Quantum Key Distribution (QKD) protocols enable the secure exchange of information between two parties, namely Alice and Bob, by sharing a symmetric secret key through a quantum channel \cite{BEN84, Pirandola_2020, ekert1991quantum}. The process involves a quantum phase where quantum information is distributed and measured, followed by classical post-processing. In this subsequent classical stage, measurement results undergo reconciliation to address any discrepancies before extracting a secret key during the Privacy Amplification phase. This research paper focuses on the Information Reconciliation phase, a critical component influencing the range and throughput of QKD systems.

Two of the most dominant algorithms used in Information Reconciliation for discrete-variable QKD systems are \textbf{Cascade} and \textbf{LDPC} codes.
Cascade \cite{pa} is one of the earliest schemes proposed for Information Reconciliation and has seen widespread use. It is based on a combination of parity exchanges and binary search to locate errors. Since its introduction, significant work has been put into the optimization of Cascade \cite{cas_intro_i, martinez_nature_comp, winnow, pedersen2013high, HU2019156, mueller2023efficient}; a compilation and comparison can be found in \cite{martinezmateo2014demystifying}. The state-of-the-art in terms of minimum information leakage is described in \cite{pacher}. High-throughput implementations of Cascade have recently been demonstrated \cite{pedersen2013high, mao2021high} even in the presence of latency on the classical communication channel.

LDPC codes \cite{ldpc_gallager} have seen major application in classical communication and have been introduced and optimized to Information Reconciliation for QKD \cite{mink2014ldpc, Elkouss_2009}. They are based on belief propagation and syndrome decoding. To overcome performance penalties that can occur with the code rate of chosen LDPC codes not matching the quantum channel, various protocols have been proposed, e.g. the rate-adaptive \cite{Elkouss_2010_rate_adaptive, rate_ldpc_proof}, Blind \cite{blind}, and symmetric Blind protocol \cite{symm_blind, vari_blind_symm}. Various improvements and optimizations \cite{untainted, indust, kasai_nb} as well as high-throughput implementations have been demonstrated \cite{mao2019high, Dixon2014}.

While both algorithms have received considerable attention in related research, their analysis often treats them as isolated components and as operating in ideal or simplified conditions. In a practical and industrial setting, numerous aspects can significantly impact the performance, such as bad QBER estimates, limitations on the classical channel (capacity and/or latency), allocated computational resources, costly verification, and quickly varying QBER. We analyze the performance of Cascade and LDPC codes in such a practical setting, evaluating and comparing the efficiency, required number of messages, and resistance to fluctuating and varying quantum bit error rates. 

This work has the following outline: In Sec. \ref{sec:IR}, the general scenario of Information Reconciliation is introduced. In Sec. \ref{sec:cas} and Sec. \ref{sec:ldpc}, Cascade and LDPC codes are introduced, respectively. In Sec. \ref{sec:qti_device}, the used QKD system is described. The results are then presented in four sections. In Sec. \ref{sec:lat}, the number of messages used by different algorithms and settings is analyzed, and the impact of latency on the classical channel is evaluated. In Sec. \ref{sec:deviation}, the impact of QBER estimates that deviate from the true QBER are analyzed. In Sec. \ref{sec:ev}, we introduce a new measure of efficiency that also respects the cost of verification and optimizes the block sizes accordingly. Finally, in Sec. \ref{sec:res_dev}, we show the performance of Cascade and LDPC codes evaluated on a live QKD system.

\section{Background}

\subsection{Information Reconciliation}\label{sec:IR}

The Information Reconciliation (IR) phase is a vital stage of the post-processing of any practical QKD system. The goal of IR in QKD is to correct any discrepancies between the proto-keys of the two (or more) parties while minimizing the information leaked to potential eavesdroppers. Before IR, Alice and Bob share some quantum information, usually in the form of qubits (Quantum bits) that are prepared and send by Alice and measured by Bob. Generally, after sifting (a measurement basis reconciliation) and parameter estimation, Alice is in possession of a binary random string of length $n$, $\mathbf{x}=(x_0,...,x_{n-1})$, $x_i = {0,1}$. Bob is in possession of his sifted measurement results which can themselves be transformed into another binary string of length n, $\mathbf{y}=(y_0,...,y_{n-1})$, $y_i = {0,1}$. The quantum channel is assumed to behave like a depolarizing channel when no eavesdropping is taking place, such that the channel can be accurately represented by a substitute channel where $\mathbf{x}$ and $\mathbf{y}$ are correlated as a binary symmetric channel. This holds as errors are typically uncorrelated and symmetric. The transition probabilities of such a channel are as follows:

\begin{equation}\label{channel}
\text{P}(y_i|x_i) = \begin{cases} 1-q & y_i=x_i,\\
q & \text{otherwise}.
\end{cases}
\end{equation}

\noindent Here, the parameter $q$ represents the channel transition probability. We refer to the symbol error rate between $\mathbf{x}$ and $\mathbf{y}$ as the quantum bit error rate (QBER). We set it equivalent to the channel parameter $q$.
In practical systems, the estimated and actual QBER may vary depending on the precision of the parameter estimation. This can have a significant impact on the performance of Information Reconciliation, as most reconciliation schemes depend in some form on the input of the estimated QBER for coding rate selection or other parameter choices. This is explored in more detail in Sec. \ref{sec:deviation}. 

In addition to qubits, Alice also sends classical messages, e.g. syndromes or parity bits, which are assumed to be error-free. From a coding perspective, this is equal to asymmetric Slepian-Wolf coding with side information at the receiver  \cite{1042242}, where the syndrome, denoted by $\mathbf{s}$, represents the compressed version of $\mathbf{x}$, and $\mathbf{y}$ is the side information. During Privacy Amplification \cite{Br_dler_2016}, any information on the final secret key gained by a potential eavesdropper during any phase of the quantum key distribution must be deducted from the final secret key length. All the information leaked during IR will be denoted by $\leakIR$. In the case of LDPC codes, assuming no rate adaptation, it can be upper-bounded by the syndrome length in bits, $\leakIR \leq m$, with $m$ being the length of the syndrome string. For the case of rate-adaption, see Sec \ref{sec:ldpc}.  For Cascade, it can be upper-bounded by the number of parity bits sent from Alice to Bob \cite{Lo_2003}, although attention has to be paid to special cases in relation to the parameter estimation phase of QKD post-processing \cite{tupkary2023using}. Using the Slepian-Wolf bound \cite{1055037}, the minimum amount of leaked information required to successfully reconcile with an arbitrarily low failure probability in the asymptotic limit of infinite length is given by the conditional entropy:

\begin{equation} \label{upper_bound_ir}
\leakIR \geq n\text{H}(X|Y).
\end{equation}

\noindent The conditional entropy $\text{H}(\cdot|\cdot)$ of the binary symmetric channel, assuming independent and identically distributed input $X$, can be expressed as
\begin{equation}
\text{H}(X|Y) = \text{H}(p) = - p\cdot \log_2(p) - (1-p)\log_2(1-p).
\end{equation}

\noindent The relative information leakage can be quantified by the efficiency $f$, given by

\begin{equation}
f = \frac{\leakIR}{n\text{H}(X|Y)}.
\end{equation}

\noindent An efficiency of $f>1$ corresponds to leaking more bits than required by the theoretical minimum of $f=1$. In practice, systems have $f>1$ due to the difficulty of designing optimal codes, finite-size effects, and the inherent trade-off between efficiency and throughput. Details on achievable information leakage, including finite-size effects, can be found in \cite{Tomamichel_2017}. A common adaption of the efficiency measure also considers the impact of frame errors, $f_{\text{FER}}$ as \cite{mao2021high, martinezmateo2014demystifying}:

\begin{equation}
    f_{\text{FER}}  = (1 - \text{FER})f + \frac{\text{FER}}{H(p)},
\end{equation}

\noindent where FER is the frame error rate. We also propose a new efficiency measure  \eqref{f_eff} in  Sec. \ref{sec:ev}  that further considers the cost of verification for different frame sizes. Apart from the $f$-efficiency notation, a different measure for the efficiency is reported in literature, the $\beta$-efficiency. The two notations are related as

\begin{equation}
    \beta (\text{H}(X) - \text{H}(X|Y)) = \text{H}(X) - f\text{H}(X|Y).
\end{equation}

\noindent The Information Reconciliation stage is followed by an Error Verification (EV) stage in which a bound on the expected probability of correctness is acquired. In this context, correctness refers to both versions of the key, i.e. Alice's and Bob's, being identical after Information Reconciliation. In practice, this is often achieved by exchanging and comparing hashes \cite{Luetkenhaus_1999, Fung_2010}.

\subsection{Cascade}\label{sec:cas}

Cascade \cite{10.1007/3-540-48285-7_35} is among the initial strategies introduced for Information Reconciliation and has been widely adopted for its simplicity and effectiveness. Subsequent studies on the original Cascade protocol have aimed to enhance its efficiency either by replacing parity exchange with error-correction techniques \cite{winnow} or by fine-tuning parameters such as top-level block sizes \cite{martinezmateo2014demystifying}. By deviating from its commonly used description, Cascade can be placed inside the framework of linear block codes \cite{pacher}. The binary Cascade protocol acting on a single frame can be summarised in the following steps, where the only expected inputs are a binary frame of length $n$ and an estimate of the QBER $\hat{q}$:

\begin{itemize}
    \item Iteration 1:
    \begin{enumerate}
    \item The binary frame is divided into non-overlapping blocks of size $k_1$, where the value of $k_1$ usually depends on the estimated QBER of the frame, e.g. the original Cascade protocol suggested a value of $k_1 = 0.73/\hat{q}$. We will refer to these blocks as ``top-level-blocks''.
    \item Alice and Bob calculate the parity of each top-level-block and share them over a noiseless and public classical channel. At all times, they track the number and length of messages they share publicly. By comparing the parities, they will detect if Bob's block has an uneven number of errors. An even number of errors will go undetected.
    \item For those top-level-blocks where a mismatch between the parity of Alice and the parity of Bob is detected, a binary search is used to detect a single error. In general, if and only if a parity mismatch between Alice and Bob occurs, a binary search can be performed on that block. The binary search consists of three steps:
    \begin{enumerate}
        \item Split the respective block in half.
        \item Calculate and exchange the parities of the two sub-blocks. In practice, it is sufficient to just share one of the parities, as each carries the full bit of information. One of the sub-blocks is guaranteed to have a mismatching parity.
        \item  If the mismatching sub-block contains only 1 bit, the error is found. Otherwise, repeat Step (a) \& (b) with the mismatching sub-block.
    \end{enumerate}
    The binary search locates the position of one error in at most $\lceil \log_2 k_1 \rceil$ steps. In Iteration 1, all top-level-blocks can be processed in parallel and all parities of a respective level of the binary search can be exchanged using a single message. We keep track of all occurring blocks and their parities. At the end of an iteration, all blocks will have 0 or an even number of errors.
    \end{enumerate}
    \item Iteration $i$:
    \begin{enumerate}
    \item Apply a permutation on the frame and divide it into new top-level-blocks of size $k_i$. The value of $k_i$ usually depends on $i$ and varies for different versions of Cascade, the original protocol suggests a doubling of the block size with each iteration, i.e. $k_i =2 k_{i-1}$. Repeat steps (ii) and (iii) as described for the first iteration on the new blocks.
    \item Cascade step: Starting from the second iteration, detected errors can be used to try to find even more errors by utilizing the name-giving idea of the Cascade protocol. First, we realize that any erroneous bit detected in iteration $i$ also takes part in different blocks created in previous iterations. When we correct the value of that bit, the parity of all blocks containing that bit will flip, such that all these blocks now have an odd number of errors. Each time we have an odd number of errors in a block, we can utilize the binary search to locate an error. We can therefore run a binary search for a block in all previous iterations. The errors we detect in those successive binary searches again take part in blocks of all other iterations and can be used to detect more errors, creating the ``cascading'' effect of the Cascade step. Note that sometimes binary search on different blocks in different iterations will point towards the same erroneous bit, requiring additional checks when parallelizing. 
    \end{enumerate}
\end{itemize}

\noindent The protocol stops after a fixed number of iterations. To the best of our knowledge, state-of-the-art in terms of efficiency with values up to $f=1.025$ is reached by a modification \cite{pacher} of the original Cascade. The changes of this modification constitute mainly of separating bits into groups of similar confidence in iteration 2. These groups are then treated separately, with different top-level-block sizes. Further and in general, choosing block sizes that are powers of two for all other iterations is shown there to be optimal. All simulations and further considerations in this work follow these modifications, it is sometimes additionally referred to as the efficient version of Cascade.

\subsection{LDPC codes} \label{sec:ldpc}

We assume some familiarity with belief propagation decoding and binary Low-density parity-check (LDPC) codes. For an overview see \cite{ldpcbook}; here we focus on the description of the rate-adaptability using the Blind protocol \cite{martinezmateo2013blind} which is a commonly used protocol for high efficiency in QKD setups. While one can achieve even better efficiency using the symmetric Blind protocol \cite{Kiktenko_2017}, we focus on the Blind protocol in this work as it maintains the asymmetric computational load between Alice and Bob, in similarity to Cascade. 
In general, an LDPC code can be represented by its parity check matrix $\mathbf{H}$, with $N$ columns and $m$ rows. In the setting of QKD, Alice has access to the original data $\mathbf{x}$, and Bob to the side information $\mathbf{y}$ which is correlated to $\mathbf{x}$ according to \eqref{channel}. Bob also receives the syndrome $\mathbf{s}$  from Alice over an error-free channel. Bob can then try to recover $\mathbf{x}$ by using methods similar to those used in classical Slepian-Wolf and LDPC coding, e.g. decoding by iterative belief propagation. LDPC codes can be designed for different coding rates, where the coding rate is defined by

\begin{equation}
    R = 1 - \frac{m}{N},
\end{equation}

\noindent where $N$ is the frame size and $m$ is the syndrome length, i.e. the number of columns and rows of $\mathbf{H}$, respectively. For perfect reconciliation, the code rate is given by the entropy of the channel as $R = 1-\text{H}(q)$, and for realistic settings one can again use the efficiency $f$ as 

\begin{equation}
    f = \frac{1-R}{\text{H}(q)} = \frac{m}{N\text{H}(q)}.
\end{equation}

\noindent In a practical scenario, only one or a few codes of respective code rates $R_i$ are available. Using a code outside its designed code rate can lead to either a decoding failure, i.e. Bob is not able to recover the correct key, or a bad efficiency, i.e. the syndrome is longer than required and too much information has been leaked to the eavesdropper \cite{blind}. This can be avoided using rate adaption techniques that allow modifying the code rate $R_i$ of a base code to match the wanted code rate $R=1-f\text{H}(\hat{q})$ \cite{Elkouss_2010_rate_adaptive}, where $\hat{q}$ is an estimate of the QBER. 
There are two main methods for this, \textbf{puncturing} for increasing the code rate and \textbf{shortening} for decreasing the code rate.
Puncturing works by adding $p$ random bits to the original message, thereby increasing the code rate. During shortening, we instead add $s$ bits whose value is publicly known. The new, adapted code rate is then given by 

\begin{equation}
    R_{i, \text{adapted}} = \frac{N-m-s}{N-p-s}.
\end{equation}

\noindent  To guarantee a fixed frame length of the corrected key, we fix the sum of punctured and shortened bits to a fixed value $d=p+s$. Given the required code rate one can then choose the correct base code rate $R_i$ and the respective number of punctured and shortened bits. This method still suffers from sensitivity to mismatched code rates, i.e. the created code rate does not match the truly required code rate due to a bad estimate of the QBER. The Blind protocol can increase resilience to this by allowing for multiple rounds of communication between Alice and Bob. We give here a summary of the original description \cite{martinezmateo2013blind}.

\noindent \textit{Preliminaries}: We have a set of LDPC codes of rates $\{R_0,...,R_l\}$ of frame size $N = n +d$, and we have knowledge about which code has the best $f_{\text{FER}}$ for each possible QBER value. Alice and Bob are in possession of their respective keys of length $n$, $\mathbf{x}$ and $\mathbf{y}$.

\begin{enumerate}
    \item Alice initially punctures all bits reserved for rate adaption, i.e. sets $p=d$.  Alice then creates an extended key $\mathbf{x}^{\prime}$ of length $N$ by filling all bit positions not reserved for rate adaption with the values of the original key $\mathbf{x}$. All bit positions reserved for rate adaption are filled with punctured bits. These positions are either random or optimized beforehand \cite{untainted}. She then calculates the syndrome $\mathbf{s} = \mathbf{H}\mathbf{x}^{\prime}$ and sends it to Bob. 
    \item Bob creates his own version of the extended key $\mathbf{y}^{\prime}$ using $\mathbf{y}$ and all information about shortened or revealed bits he has. He tries to recover $\mathbf{x}^{\prime}$, if this fails after a given number of iterations of belief propagation, he asks Alice to turn punctured bits into shortened bits or reveal key bits if necessary. He asks for $v = \lceil n \cdot (0.028 - 0.02R) \cdot \alpha \rceil$ bits \cite{Kiktenko_2017}, where $\alpha$ is a step size that can be optimized. He chooses those bits for which he has the lowest confidence, i.e. the smallest absolute value of the posterior.  This is a slight deviation from the original description \cite{martinezmateo2013blind} introduced in \cite{Kiktenko_2017}.
    \item Alice receives Bob's request and sends him the requested bit values. Steps (ii) and (iii) are repeated until either Bob declares a successful decoding or all bits of the key are revealed. In a practical implementation, one can already stop and declare failure once the amount of leaked information no longer allows for secret key extraction.
    \item After decoding, Bob and Alice recover the keys without the rate modulated bits by selecting only those bit positions that belonged to the original input, and recover $\mathbf{x}$ and $\mathbf{\hat{x}}$ from $\mathbf{x}^\prime$ and $\mathbf{y}^\prime$, respectively, where $\mathbf{\hat{x}}$ is Bob's guess of $\mathbf{x}$. 
\end{enumerate}

\noindent The extended frame is then passed to the next module, e.g. Error Verification, and the amount of leaked information $\leakIR$ is calculated using the effective rate $R_{i, \text{adapted}}$ of the last try \cite{rate_ldpc_proof} and the number of revealed bits $k_{\text{revealed}}$,

\begin{equation}
    \leakIR = \lceil n(1-R_{i, \text{adapted}}) \rceil + k_{\text{revealed}}.
\end{equation}

\subsection{The QKD system} \label{sec:qti_device}

The deployed QKD system is provided by Quantum Telecommunications Italy (QTI). It is composed of the transmitter and receiver equipment, here named Alice and Bob, respectively. The quantum link consists of a low-loss optical fiber and has a channel loss of 10 dB. The classical channel, through which the post-processing is performed, is a 40km dedicated fiber. The QKD protocol implemented is a prepare\&measure 2-dimensional three-state BB84 protocol, using the 1-decoy state method \cite{rusca2018finite}.

The three-state protocol uses two different quantum states in the code-basis (Z), and one state in the test basis (X), the two bases are mutually unbiased. The three possible quantum states are prepared by Alice and encoded using a time-bin encoding scheme \cite{Bacco2019FieldTO, ribezzo2023deploying}. Given two temporal intervals, 
the states of the Z basis
are defined by the time of occupation of the signal, first and second, respectively. 
The state prepared in the X basis
is a superposition of the Z-basis states
, with zero relative phase between the two pulses. 

\begin{figure}
    \centering
    \includegraphics[width=0.9\textwidth]{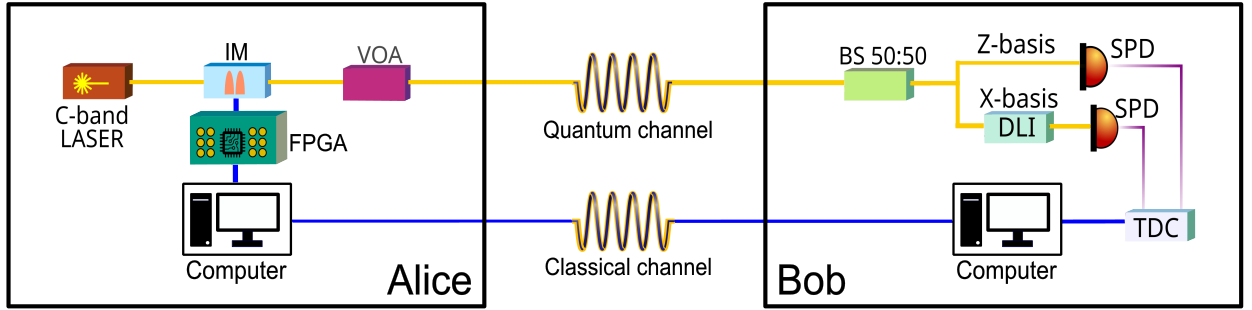}
    \caption{Schematics of the experimental setup. FPGA:  field programmable gate array, IM: intensity modulators, VOA: variable fiber optical attenuator, BS: beam-splitter, DLI: delay line interferometer, TDC: time to digital converter, SPD: single-photon detector.}
    \label{figure: setup}
\end{figure}

The pulses encoding the qubits are created by carving a continuous-wave C-band laser via two intensity modulators driven by a 595 MHz signal. The pulses are then attenuated to reach the single-photon level. In particular, two different intensities are prepared to implement the 1-decoy method. 
At the receiver side, the quantum states are measured by two InGaAs/InP single-photon detectors (SPD). A 50:50 beam splitter is used to re-direct the incoming qubits to the two mutually unbiased bases (Z and X). For the Z-basis, a SPD measures the time of arrival of the photons. To extract information of the states prepared in the X basis, an unbalanced interferometer makes the phase measurement possible.
The schematics of the experimental setup can be seen in Figure \ref{figure: setup}.
After the acquisition stage, Alice and Bob respectively own a set of quantum measurements that must be properly sifted. They proceed to a second stage of the protocol, the so-called post-processing, communicating through an authenticated classical channel. The post-processing includes sifting, parameter estimation, error correction and verification, and Privacy Amplification. It extracts a correct and secret shared key from the raw measurements. 

The secure key is distilled from the Z-basis detections, while the X-basis detections are used for the security analysis. 
From the perspective of studying the performances of the algorithm used for  error correction, the secure key length $\text{skl}$ can be expressed as 

\begin{equation*}
    \text{skl}= \text{skl}_\text{ind} - \leakIR - \leakEV,
    \label{eq:sec length}
\end{equation*}
where the first term $ \text{skl}_\text{ind}$ is system-dependent but independent of the chosen algorithm (Cascade or LDPC/Blind Protocol). The terms $ \leakIR$ and $\leakEV$ are bits of information 
leaked to a malicious adversary during IR and EV. Finally, the secure key rate is given by the fraction of $\text{skl}$ over the time of the QKD round, i.e. the overall time of the key generation.  For the experiment, i.e., Sec \ref{sec:res_dev}, the hardware settings of the system have been changed, 
resulting in an increasing quantum bit error rate with time, as this is beneficial for the analysis of the error correction algorithm's behavior.

\section{Results}

In this section, we analyze different measures of performance for both Cascade and the Blind protocol with LDPC codes in relevant practical and industrial settings and compare them, first in simulations and later on an industrial QKD system. The simulations conducted in Sec. \ref{sec:lat}, Sec. \ref{sec:deviation}, and Sec. \ref{sec:ev} are based on data collected from an industrial QKD system. In Sec. \ref{sec:res_dev}, we evaluate the performance of LDPC codes and Cascade on the system in a live test. The system itself is described in Sec. \ref{sec:qti_device}.

\subsection{Latency and the number of exchanged messages} \label{sec:lat}

The most significant figures of merit for any QKD Information Reconciliation system are its information leakage (related to the efficiency $f$) and its throughput, i.e., how much information can be processed per time unit. These two figures are often related in a trade-off involving different parameters of a particular error correction algorithm, i.e. a better efficiency results in a lower throughput. From a perspective of optimizing the secret key rate of a system, one would therefore like to choose an algorithm and parameters that can handle the maximum required throughput while having the best efficiency. While both, Cascade and LDPC codes, have a variety of versions that can be used for QKD, we restrict ourselves in this work to two specific versions, Cascade as by Pacher \cite{pacher} and the Blind protocol for LDPC codes \cite{martinezmateo2013blind}. The efficiency of the LDPC codes could be further increased by instead using the symmetric Blind protocol \cite{Kiktenko_2017}, at the penalty of additionally using significant computational powers at Alice. To the best of our knowledge, the used version of Cascade obtains the best efficiency out of all Cascade implementations.

Cascade is a communication-heavy protocol and concerns about throughput limitations have been voiced in the years after its introduction \cite{Martinez-Mateo2013, winnow, jouguet2013high}.  Recently, it has been shown that a  high throughput can be achieved nevertheless by using sophisticated software implementations \cite{pedersen2013high, mao2021high}. LDPC codes have been shown to achieve similar throughput for QKD \cite{mao2019high}. 
As these papers show, the throughput of either algorithm is very dependent on the software implementation and specific parameter choices, in particular for LDPC codes. We therefore focus our following analysis on characteristics that are dominated by the algorithm design itself, including efficiency, latency, and the expected number of messages required between the parties. 

\begin{figure}
    \centering
    \includegraphics[width=\linewidth]{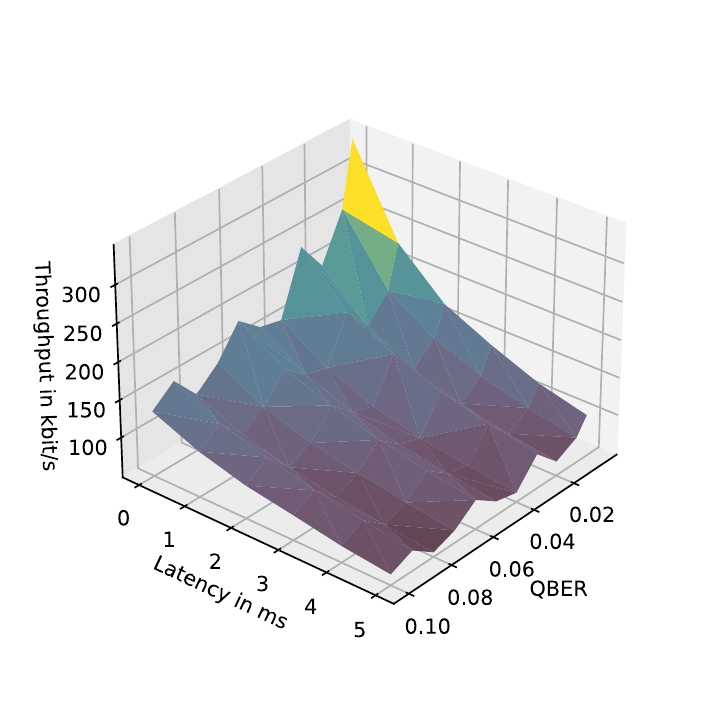}
    \caption{Throughput behavior of the Cascade implementation. The algorithm is a parallelized version of the description in \cite{pacher}. The latency is given for a one-way connection. The frame size is chosen as $2^{16}$. The results shown are simulated assuming a binary symmetric channel.}
    \label{fig:thru-cas}
\end{figure}

We evaluate the throughput of Cascade for a block-size of $2^{16}=65536$ bits (it has been shown in \cite{pacher} that further increasing the block-size does not substantially increase the efficiency while larger frames can result in penalties from a computational point of view \cite{mao2021high}) by varying the latency and QBER on the channel. The results can be seen in Fig. \ref{fig:thru-cas}. Despite the relatively high number of messages required for the efficient version of Cascade (between 400 and 700 for a frame of size $2^{16}$), the decline in throughput is only around $60\%$ going from 1ms to a 5ms latency on a serial execution. A 5ms delay corresponds approximately to a 1000 km distance in optical fiber, a distance at which the expected raw key rate (and therefore the required throughput) drops by several magnitudes of order for contemporary systems that do not have access to quantum repeaters. Current research systems that can reach these distances have key rates below one bit per second \cite{Liu_2023}. 

\begin{figure}
    \centering
    \includegraphics[width=\linewidth]{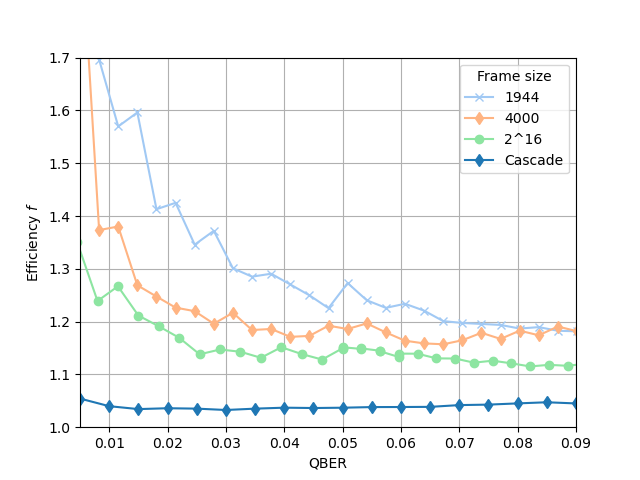}
    \caption{Efficiency $f$ for LDPC codes of different frame lengths and QBER values. The frame length of Cascade is the same as the longest evaluated LDPC codes, i.e. $2^{16}=65536$.}
    \label{fig:eff-all}
\end{figure}

\begin{figure}
    \centering
    \includegraphics[width=\linewidth]{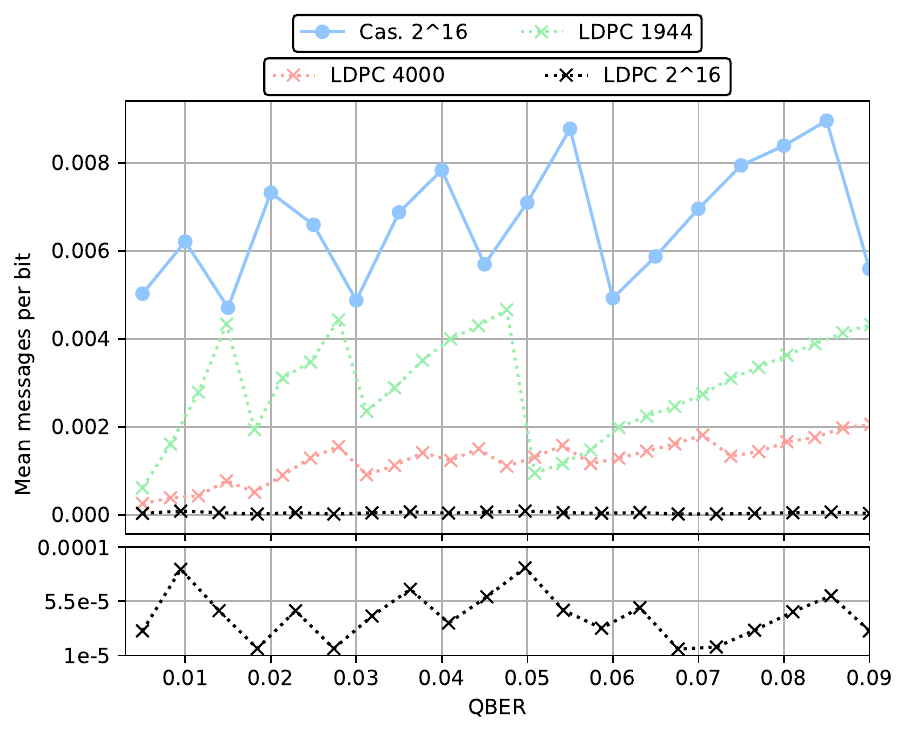}
    \caption{Mean number of messages sent per reconciled bit using Cascade and LDPC codes for different frame sizes.}
    \label{fig:msgs}
\end{figure}

The throughput and efficiency of the Blind scheme have a strong dependency on the size of the chosen codes, various decoder parameters, and the step-size of the Blind protocol, giving significantly more space for variation by design than Cascade. We evaluate the performance of 3 different sets of LDPC codes of size $n_1=1944$, $n_2=4000$, and $n_3 = 2^{16}$. The code set of size $n_1$ contains 4 standard industry LDPC codes \cite{5307322, Kiktenko_2017} (used in classical communication), the codes of size $n_2$ and $n_3$ are constructed using the optimized degree distributions from \cite{Elkouss_2009} and the PEG algorithm \cite{1377521}. Each set contains 10 codes, covering different code rates. Their evaluated efficiency can be seen in Fig. \ref{fig:eff-all}, together with the well known efficiency of Cascade. We use the Sum-Product algorithm (SPA) decoder \cite{spa} with 50 decoding iterations and a step size $\alpha=1$ for the Blind protocol. We also visualize the number of messages exchanged for different frame-sizes in Fig. \ref{fig:msgs}, where the mean number of messages required per corrected bit is shown. One can see that for smaller frame-sizes, the number of required messages is not significantly smaller than for Cascade,  and only at larger frame-sizes does LDPC show a clear advantage in the number of messages sent. Due to this and reasons specified further in Sec. \ref{sec:disc}, we focus on the larger block-size in the following analysis.



\subsection{Performance for deviating QBER-estimates} \label{sec:deviation}

It is common practice in QKD security proofs that a parameter estimation phase is performed before the Information Reconciliation step \cite{renner2006security, Scarani_2009}. During parameter estimation, an estimate $\hat{q}$ of the true QBER $q$ is established which is used to select a fitting code rate or other parameters that are relevant for Information Reconciliation. The bits used for estimation are revealed and can therefore no longer be used for secret key extraction. In various practical systems, i.e. the system analyzed in this work or other systems aimed at industrial application \cite{Kiktenko_2016}, a different approach is used. After a successful error correction for the $i$-th key, Bob is in possession of both the original measurement results as well as the corrected key, allowing him to calculate the true error rate $q_i$ by comparing both. In a continuously running system, the true error rate of the current key can then be used as an estimate for the next iteration, $\hat{q}_{i+1} = q_i$. Given this setup, we investigate two main aspects in the following analysis:

\begin{itemize}
    \item Good block lengths for estimating the QBER.
    \item The impact of mismatch between $q_i$ and $\hat{q}_i$ on the performance of Cascade and LDPC codes.
\end{itemize}

\begin{figure}
    \centering
    \includegraphics[width=\linewidth]{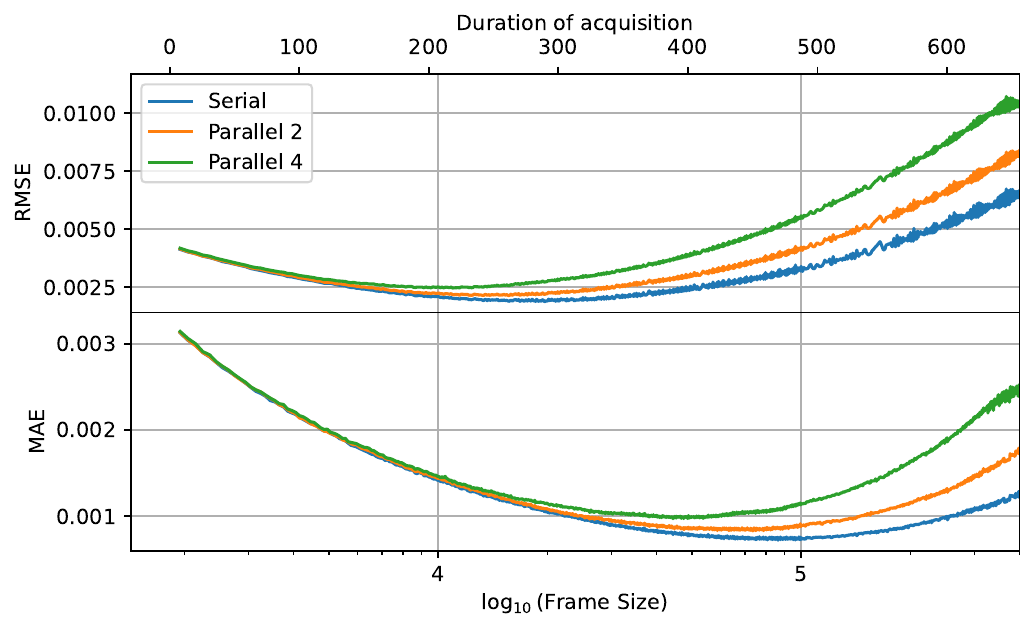}
    \caption{Quality of QBER estimation for different frame sizes used. RMSE refers to the root mean square error and MAE refers to the mean absolute error. Serial and Parallel denote a directly serial processing and parallel processing of 2 and 4 frames, respectively.}
    \label{fig:qber_mae}
\end{figure}

\noindent We can choose different frame sizes on which to evaluate $q_i$ by using either the complete key, a subdivision, or by combining multiple keys. We use data collected from the QKD system which we divide into blocks of size $k$ on which the true QBER $q_i$ is evaluated. We then calculate different measures on the deviation between the estimated  $\hat{q}$ and true QBER $q$ of all blocks. The results are shown in Fig. \ref{fig:qber_mae}, where different measures are shown: MAE refers to the mean absolute error and RMSE is the root mean square error, i.e. a 1-norm and 2-norm, respectively. As one would expect, there is an optimum for medium-sized blocks where the block size is large enough to allow for precise sampling but the acquisition time required is short enough to represent the time scale of relevant error sources. This behavior is highly dependent on the actual system used as it is strongly influenced by the acquisition rate and the time scale of possible error sources but the method itself can be used for any system. The results for the tested system align with the order of block-size choices we made for the error correction algorithms, i.e., it is justified to simply use the same frames used for Information Reconciliation for estimating the QBER. 

Many works showing high throughputs rely on heavy parallelization of the error correction algorithms to reach high throughput \cite{mao2019high, mao2021high}. For a parallelized system, the QBER information of previous frames is not yet available at the required time due to lag and one has to fall back on earlier estimates, e.g., $\hat{q}_{i+1}=\hat{q}_{i+2}=q_i $ for 2 parallel frames. In Fig. \ref{fig:qber_mae}, we show the impact of running 2 or 4 different sequential frames in parallel on the quality of the QBER estimate. The acquisition rate is held constant. The results suggest that it is advantageous to stay serial if possible and choose shorter block-sizes for estimation when running multiple frames in parallel.
We emphasize that finding the truly optimal block length, with respect to the secret key rate, requires additionally considering the performance of the used error correction with respect to small deviations and outliers (observe that the 1 and 2 norms have different minima), and then optimize the efficiency accordingly.

\begin{figure}
    \centering
    \includegraphics[width=\linewidth]{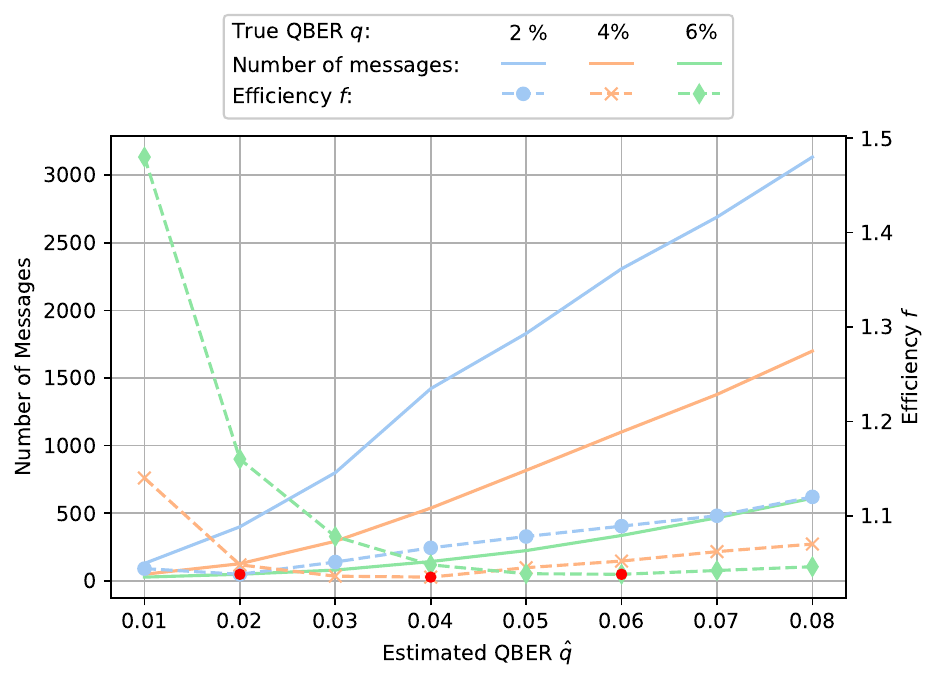}
    \caption{Number of exchanged messages and the efficiency $f$ for different QBER deviations for Cascade. Estimated QBER denotes the QBER that has been given to the reconciliation algorithm as input, whereas the true QBER refers to the QBER of the actual data.}
    \label{fig:bad_qber_cas}
\end{figure}

\begin{figure}
    \centering
    \includegraphics[width=\linewidth]{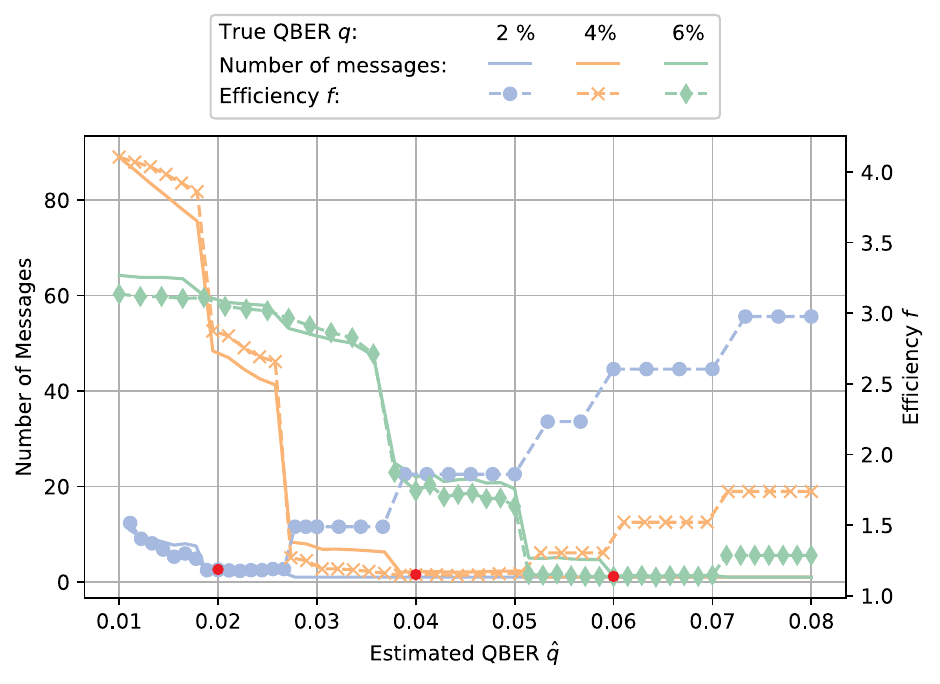}
    \caption{Number of exchanged messages and the efficiency  $f$ for different QBER deviations for the Blind protocol. Estimated QBER denotes the QBER that has been given to the reconciliation algorithm as input, whereas the true QBER refers to the QBER of the actual data.}
    \label{fig:bad_qber_ldpc}
\end{figure}

The performance of Cascade and the Blind protocol for LDPC codes under a mismatch between the estimated QBER and the true QBER can be seen in Fig. \ref{fig:bad_qber_cas} and Fig. \ref{fig:bad_qber_ldpc}, respectively. The true QBER is fixed, while the estimated QBER that is given as input to the algorithm is varied. The graphs show the behavior for 3 different true QBER values of $2\%$, $4\%$, and $6\%$. The efficiency and the number of messages required are then evaluated. 
The efficiency behavior for the original proposal of Cascade has, to a lesser extent, also been studied in \cite{martinezmateo2014demystifying}.  A  more stable behavior can be observed for the version of Cascade used in this work, where the efficiency varies little, i.e. for a mismatch of less than $3\%$ the efficiency is still below 1.1, $f<1.1$. It is often advantageous to overestimate the QBER, as an underestimation has a higher penalty. Bad estimates can have a high impact on the number of messages sent, especially for a low true QBER. For Cascade, overestimating the QBER increases the required number of messages. Overestimation by half a percent at a true QBER of $2\%$ results in an increase in the number of messages sent by more than $50\%$ already. If one underestimates the QBER, the number of messages is instead reduced. 

For LDPC codes, to understand the step pattern visible in Fig. \ref{fig:bad_qber_ldpc}, one has to recall that the Blind protocol always starts out with all possible bits punctured after a code rate is chosen. It is therefore ''blind" to QBER mismatches as long as the mismatch does not result in choosing a code of a different base rate, hence the name. The penalty in efficiency is higher for low true QBER values, and, similar to Cascade, it is better to overestimate than to underestimate the QBER. Large deviations result in efficiencies that are much worse than the same case for Cascade. The shape of this curve can be influenced by choosing a different number of codes in a trade-off between stability and base-efficiency. The number of messages goes down to a single message for overestimating the QBER and increases when underestimating the QBER. We found no significant impact on the frame error rate, i.e. the frame error rate of Cascade stayed below $0.003$ (evaluated on 1000 samples) for both perfect match and any other tested mismatch between $q$ and $\hat{q}$. There were no frame errors for the Blind protocol.

\subsection{Frame error rate and verification cost} \label{sec:ev}

\begin{figure}
    \centering
    \includegraphics[width=\linewidth]{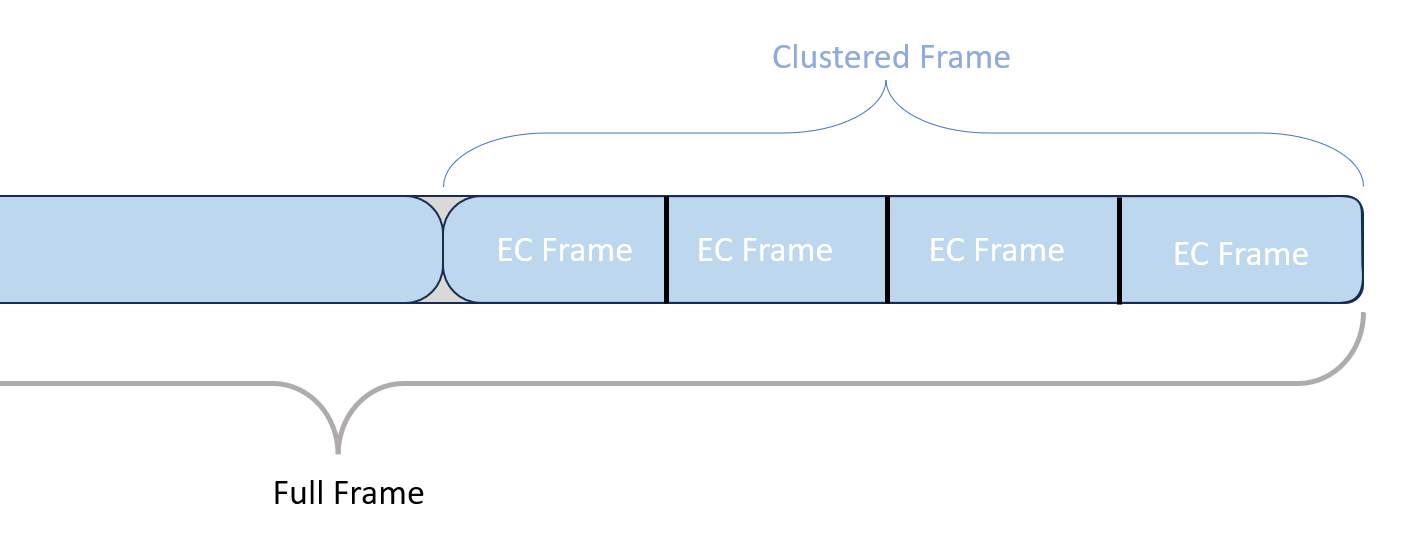}
    \caption{Schematic of the different frame notations used. The \textbf{Full Frame} refers to the frame that has an associated estimated phase error rate from the parameter estimation phase of QKD and is further used as the input frame for Privacy Amplification. \textbf{Clustered Frame} refers to the frame size that is used for Error Verification while \textbf{EC Frame} refers to the sub-frames used for the error correction.}
    \label{fig:frames_cluster}
\end{figure}

\begin{figure}[htbp]
    \centering
    \includegraphics[width=\linewidth]{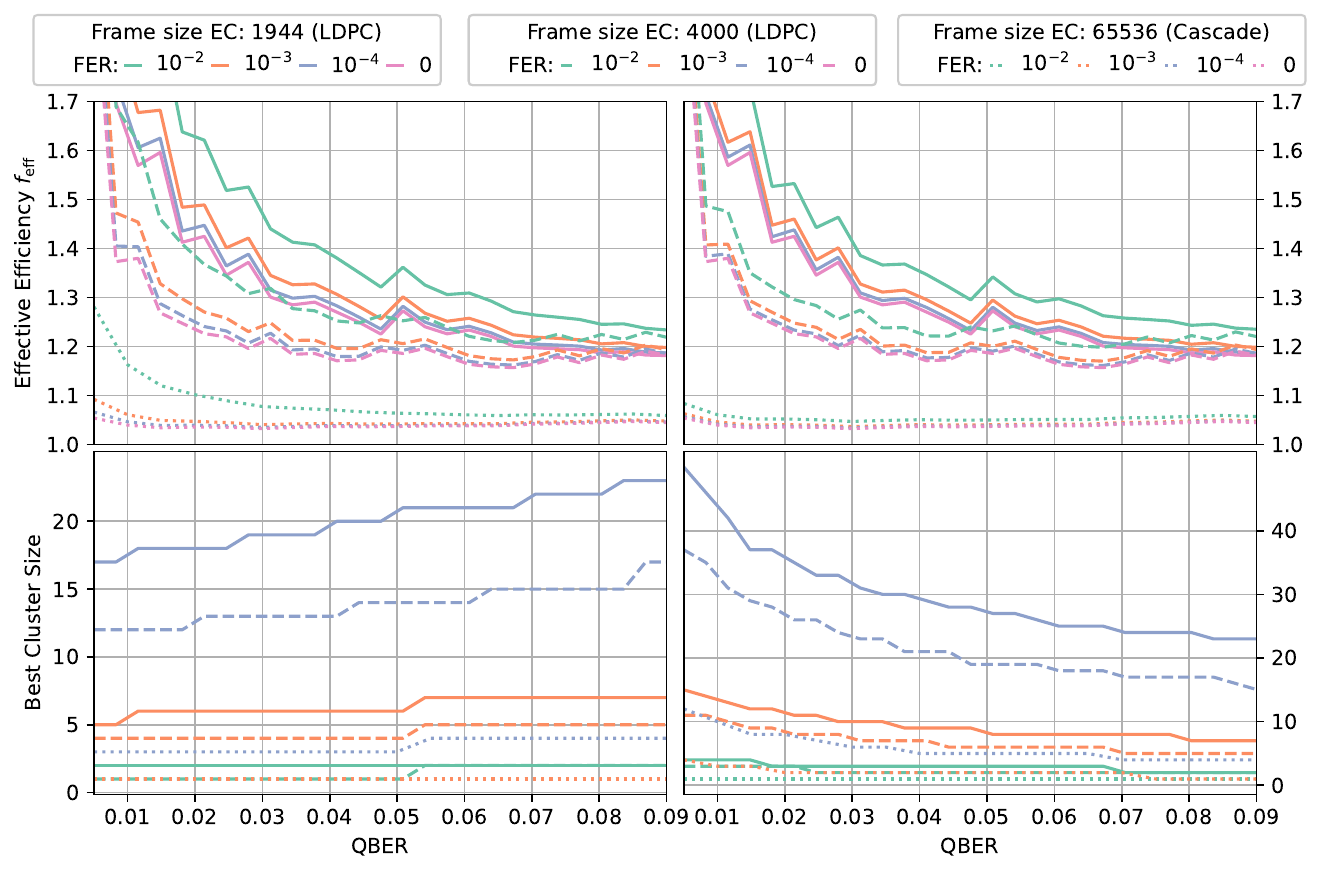}
    \caption{The performance of different block-sizes including the FER and the cost of Error Verification for different FER and QBER. Top left: Effective efficiency using the optimized cluster size. Bottom left: Optimized cluster frame size. Top right: Effective efficiency using the optimized cluster size while allowing for a repeat request for failed frames. Bottom right: Optimized cluster frame sizes while allowing for a repeat request for failed frames.}
    \label{fig:full}
\end{figure}

In addition to the leakage from IR, EV is leaking further information to the eavesdropper. During EV, the correctness of the output of IR is verified by exchanging short (public) hashes. The cost of verification should therefore be considered when optimizing. Privacy Amplification is performed on a frame length (full frame) that can be significantly larger than the frames used for error correction (EC frame) to avoid finite-size penalities \cite{Scarani_2009, 10.1007/3-540-48285-7_35, Chaiwongkhot_2017, Sidhu2022}. 
In the case of reconciliation without frame errors, a single verification of the full frame is sufficient to ensure the correctness of the final key. However, in practice, a non-zero Frame Error Rate (FER) might necessitate additional verification on the smaller frames used in error correction, to mitigate the impact of larger failed frames. On the other hand, verifying all small EC frames can be costly due to the potential leakage of verification tags. A more efficient approach is to cluster multiple corrected EC frames and perform a single verification on the combined, longer frame (clustered frame). This approach reduces verification costs while maintaining a low failure probability. For an overview of the different frames used in this process, refer to Fig. \ref{fig:frames_cluster}.
We propose the following effective efficiency $f_{\text{eff}}$ to evaluate performance in a combined IR and EV framework:

\begin{equation} \label{f_eff}
    f_{\text{eff}} = (1-\text{FER}_{\text{cluster}}) f + \frac{\text{FER}_{\text{cluster}} + \text{P}_{\text{Collision}}}{\text{H}(q)} + \frac{t}{n \text{H}(q)},
\end{equation}

\noindent where $f$ is the efficiency evaluated as $f= \leakIR/(n\text{H}(q))$, $\text{H}(\cdot)$ is binary entropy, $t$ is the number of bits used for the tags of EV, $n$ is the frame size of error correction, $\text{P}_{\text{Collision}}$ is the probability of failure of the EV method, and $\text{FER}_{\text{cluster}}$ is the chance for a clustered frame to be incorrect, 

\begin{equation}
    \text{FER}_{\text{cluster}} = 1 - (1 - \text{FER})^{k},
\end{equation}

\noindent where $k$ is the number of blocks used to create a cluster. The different terms in \eqref{f_eff} can be understood rather intuitively. The first term refers to the usual cost of error correction that has to be paid for all frames that are corrected successfully. The second term includes the chance of a clustered frame to fail, and an upper bound on the verification not detecting an erroneous frame, while the last term includes the cost of verification. The optimal number of corrected frames to cluster together for Error Verification can be calculated using \eqref{f_eff} and is visible in Fig. \ref{fig:full} for different fixed FER and block-sizes. We evaluated block-sizes 1944, 4000, and $2^{16}$, where the respective efficiency curves of 1944 LDPC, 4000 LDPC, and Cascade (see Fig. \ref{fig:eff-all}) have been used. The best effective efficiency $f_{\text{eff}}$ reached by those values can also be seen in Fig. \ref{fig:full} with $t=50$ and $\text{P}_{\text{Collision}} = 1^{-10}$. One can observe a significant penalty, especially for high FER, short block-lengths, and low QBER. The high impact for low QBER can be understood by considering the relatively high cost of a failed frame compared to the low cost of regular leakage at those QBER values. Per design, the Blind protocol does not allow for frame errors as it will always converge by revealing more and more symbols; only a syndrome error can lead to a frame error. The shown behavior is therefore mainly relevant in case one limits the number of tries the Blind protocol is allowed to use, or when using the rate-adaptive scheme.
This behavior can be improved by allowing for repeated tries on failed frames, i.e. Alice and Bob repeat the error correction for all error correction frames inside a cluster frame for which the verification detected an error. Fig. \ref{fig:full} further shows the best effective efficiency and cluster size while allowing for a single repeat request per clustered frame. The performance penalty is reduced, especially for low QBER, and cluster sizes are significantly bigger.

\subsection{Performance on live system}\label{sec:res_dev}

\begin{figure}
    \centering
    \includegraphics[width=\linewidth, trim={1cm 2cm 1cm 3cm},clip]{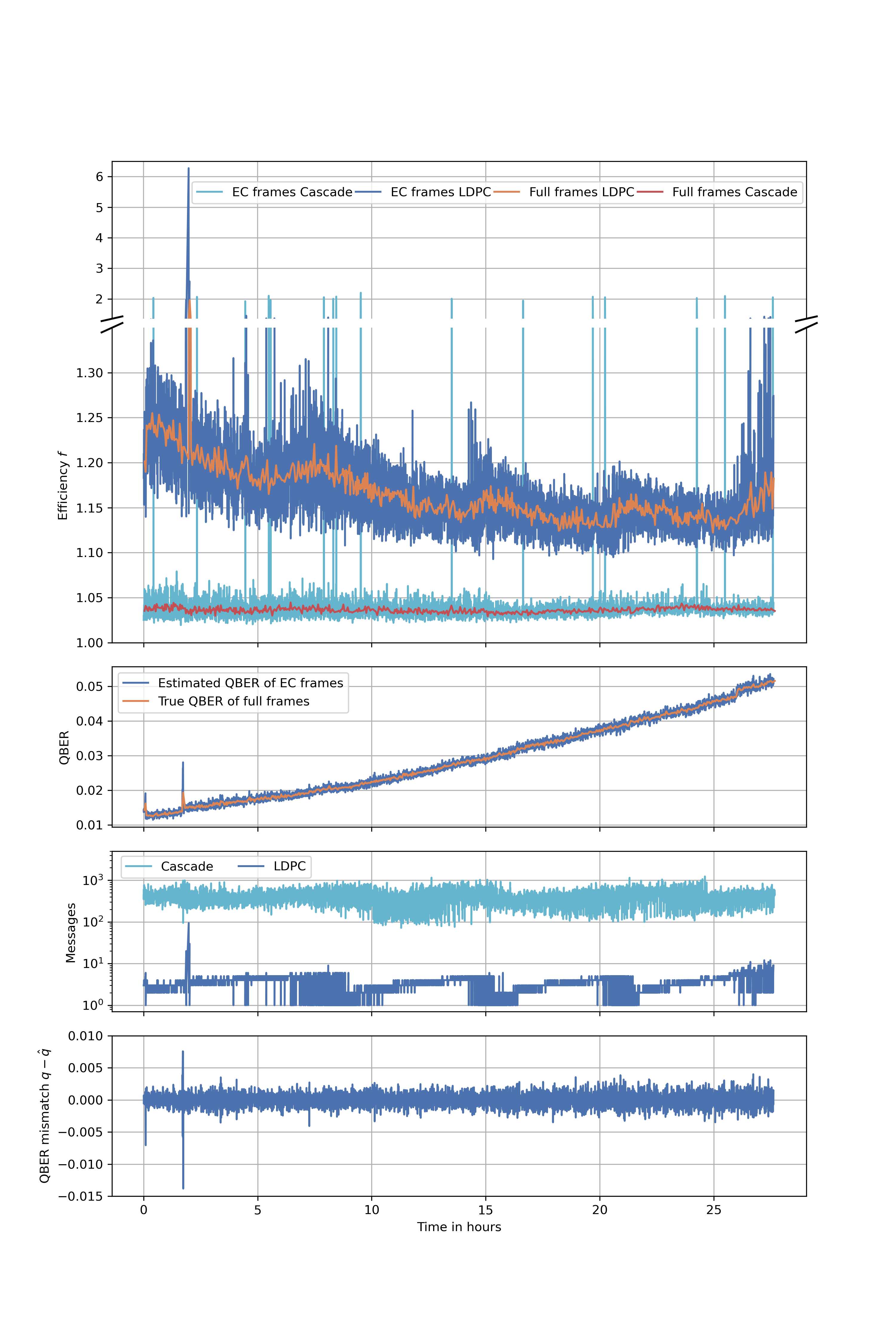}
    \caption{Different figures of merit evaluated on an industrial QKD system using Cascade and the Blind protocol with LDPC codes.}
    \label{fig:device}
\end{figure}

We evaluated the performance of both Cascade and the Blind protocol on the industrial QKD system described in Sec. \ref{sec:qti_device}. Data was taken continuously for around 27 hours and processed in real-time on the system, see Fig. \ref{fig:device}. During the experiment, the throughput of the raw data acquisition stage, i.e. frame size over time of acquisition and sifting, was around 6.7kbit/s. The topmost graph shows the efficiency for the frames of size $2^{16}$, i.e. the frame-size used for error correction, and for the full frames, i.e. the frames on which the phase error rate estimation is taken and that will later be used for extracting secret key during Privacy Amplification. On average, with dependency on the acquisition rate, $8.5$ EC frames constitute 1 full frame. The spikes of the EC frames of Cascade are due to failed frames and consequential repetition of the algorithm and doubling of the leaked information. The single spike for LDPC can be attributed to the Blind protocol requiring a substantial number of revealed symbols to decode, as the QBER has been underestimated for this frame. The Blind protocol has no frame errors by design apart from syndrome errors which we did not encounter in this experiment. 

For the LDPC codes, the $2^{16}$ sized code and decoder described in Sec. \ref{sec:lat} is used. 
As a stress test, the temperature control of the system has been turned off in this experiment to sweep the QBER and acquire general performance results. In the lowermost graph, one can observe that the QBER estimation by using the true QBER of the previous frame is remarkably good for this system; the deviation being below $0.5\%$ almost all of the time. This results in a very minor penalty on the efficiency for both approaches. The mean efficiencies of Cascade and LDPC are $f_{\text{Cascade}}=1.036$ and $f_{\text{LDPC}}=1.166$, respectively. The mean number of messages are $446$ and $3.14$.

\section{Discussion} \label{sec:disc}

A core aspect of this work is to compare the performance of Cascade and LDPC codes for IR on a practical QKD system. Both methods, but especially LDPC codes, have a wide range of tweaks and trade-offs that can significantly influence performance. In the case of Cascade, this is mainly restricted to different choices of block- and frame-sizes  \cite{martinezmateo2014demystifying, HU2019156}, and minor algorithmic changes \cite{pacher, winnow}. In the case of LDPC codes the choices are arguably plentier, some examples being the code length \& construction \cite{Elkouss_2009, IPEG1}, protocol (e.g. Blind, symmetric Blind, rate-adaptive \cite{blind, symm_blind, rate_ldpc_proof}), decoder \cite{spa, min_sum, emin_sum}.

Nevertheless, to the best of our knowledge, there are no LDPC codes that can reach efficiencies as low as Cascade. Assuming both algorithms can achieve the required throughput of the system, Cascade should be the Information Reconciliation of choice if one wants to optimize the efficiency $f$ or $f_{\text{eff}}$ (and by that the secret key rate). This performance is achieved at the cost of requiring many messages, i.e. about 2 orders of magnitude more than an LDPC code of similar size, see Fig. \ref{fig:msgs}. The required messages for LDPC codes can further be reduced to a single message by using a direct rate-adaptive approach instead of the Blind protocol, at the cost of reduced efficiency.

While our results only show a minor penalty for high latency on the classical communication channel on the throughput of Cascade (in accordance with \cite{pedersen2013high, mao2021high}) one can imagine issues that arise one QKD systems that are not fiber-based and have a high latency and/or low bandwidth, e.g. satellite-based QKD \cite{satellite_qkd}. On average, Cascade used more than twice the data usage on the classical channel per corrected frame compared to LDPC codes in our implementation. Using a simple estimation, we consider that each leaked bit in Cascade requires the description of a block. In the Blind protocol, only the syndrome plus information about shortened/revealed bits is required. We, therefore, expect the difference to be between 4 and 10 times depending on the QBER if one were to minimize the sent information on the classical channel, only considering the actual payload. 

\begin{figure}
    \centering
    \includegraphics[width=\linewidth]{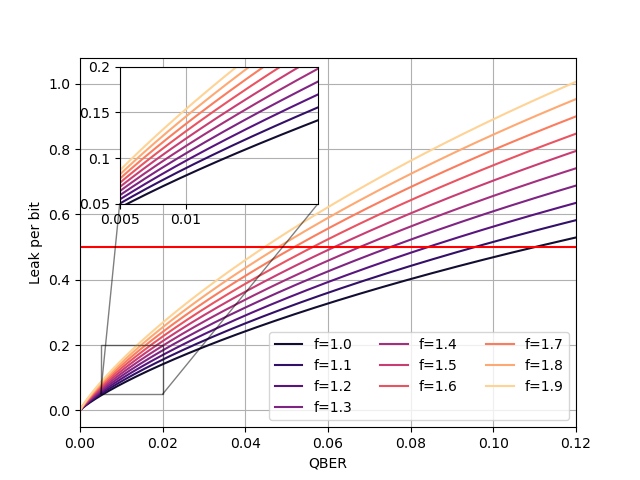}
    \caption{The leaked information per processed bit for different efficiencies in an asymptotic setting. The horizontal red line marks where even in the best asymptotic case secret key is no longer extractable.}
    \label{fig:impact}
\end{figure}

Cascade has a significantly better performance for the efficiency penalty on high deviations of the QBER estimate input, with efficiencies below 1.2 for mismatches up to $4\%$ while the Blind protocol in contrast can rise up to 3 for a similar range. Both perform well for minor deviations, the Blind protocol even being agnostic to them as long as the estimate does not change the choice of the base code. Both approaches have a significant asymmetry in the computational resources required and in the outgoing communication between both parties. Using puncturing further requires both Bob and Alice to have access to a true random number generator when using LDPC codes \cite{symm_blind}.
Despite common usage in prior works \cite{symm_blind, indu_short}, our results suggest against the usage of short LDPC codes. Their efficiency is lacking compared to Cascade and longer LDPC codes while still requiring a substantial amount of communication when using the Blind protocol. Further, the impact of verification cost is rather high when frame errors appear, i.e. when one uses the rate-adaptive protocol instead of Blind to reduce communication costs, even while using optimal clustering.
While the efficiency of LDPC codes looks rather high for low QBER values compared to the performance of Cascade, one should keep in mind the actual impact of the efficiency on the secret key rate, i.e. the ultimate performance figure of any QKD system. Fig. \ref{fig:impact} shows the ratio of leaked information per bit for different QBER and efficiencies. It is visible that the impact of sub-optimal efficiencies is rather small for low QBER values and rises with worse QBER, e.g. going from $f=1.0$ to $f=1.4$ results in a $3\%$ increase in raw key consumption due to leakage during Information Reconciliation at $1\%$ QBER but for the same efficiency jump results in $16\%$ more raw key consumption at a QBER of $8\%$.

\section{Conclusion}

We analyzed different relevant aspects of Information Reconciliation for the approaches of Cascade and LDPC codes using the Blind protocol with an emphasis on use in industrial QKD systems. The efficiency and number of required communication rounds have been evaluated for different scenarios and varying quality of the QBER estimate. We have shown that Cascade is especially stable even under large deviations ($f<1.1$ for deviations below $3\%$) while being penalized with a large number of messages required if the QBER is underestimated (more than 3000 messages per frame). LDPC codes follow a similar pattern for the number of messages but stay below 100 messages per frame. They are almost agnostic to small QBER deviations but show significant penalties in efficiency for overestimated QBER values ($f>2.5$). We showed that there is an optimal block size to use for the QBER estimation, it is of similar order as the frame size used for Information Reconciliation  ($\approx 50000$) for the exemplary system we evaluated. Further, we introduced a novel optimization of the efficiency that also considers the cost of Error Verification. We have shown that it is beneficial to cluster together smaller blocks of the Information Reconciliation and run a joint Error Verification to reduce the leaked information. This process is especially impactful if one allows for at least a single repeat request if a frame error is detected by Error Verification, with more than $40$ frames clustered together being optimal in some cases. The simulation results have further been confirmed on an industrial QKD system.

\ack

This work was funded by the Center of Excellence SPOC (ref DNRF123), the European Union (ERC, QOMUNE, 101077917), by the Project EQUO (European QUantum ecOsystems) which is funded by the European Commission in the Digital Europe Programme under the grant agreement No 101091561, the Project SERICS (PE00000014) under the MUR National Recovery and Resilience Plan funded by the European Union - NextGenerationEU, the Project QuONTENT under the Progetti di Ricerca, CNR program funded by the Consiglio Nazionale delle Ricerche (CNR) and by the European Union, the Project QUID (Quantum Italy Deployment) funded by the European Commission in the Digital Europe Programme under the grant agreement No 101091408.

\section*{Competing Interests}
The authors declare no competing interests.

\section*{Data Availability}
All data used in this work are available from the corresponding author upon reasonable request.


\section*{References}
\bibliography{main}

\end{document}